\documentclass[Royal]{sagej}

\usepackage{moreverb,url}
\usepackage[colorlinks,bookmarksopen,bookmarksnumbered,citecolor=blue,urlcolor=blue]{hyperref}
\usepackage{tikz-opm}
\usepackage{amsmath, amsthm, amscd, amsfonts,graphicx}
\theoremstyle{definition}

\newcommand\BibTeX{{\rmfamily B\kern-.05em \textsc{i\kern-.025em b}\kern-.08em
T\kern-.1667em\lower.7ex\hbox{E}\kern-.125emX}}

\begin{document}


\title{The logic of behavior of atoms}

\author{Madjid Eshaghi Gordji\affilnum{1}, Gholamreza Askari\affilnum{1} and Alireza Tavakoli Targhi\affilnum{2}}

\affiliation{\affilnum{1}Department of Mathematics, Semnan University P.O. Box 35195-363, Semnan, Iran \\
\affilnum{2}ISPR Lab., Department of Mathematics, Shahid Beheshti University, Tehran, Iran.}


\email{meshaghi@semnan.ac.ir; g.askari@semnan.ac.ir; a\_tavakoli@sbu.ac.ir }

\begin{abstract}
Understanding of the atomic structures and ways in which the atoms interacting is critical to the understanding of chemistry. In this paper applied the concepts of game theory in explaining reactions between elements of the periodic table. The findings in this study suggest that the coordination and anti coordination, also cooperation and non-cooperation between atoms and molecules lead to bonding formation according to the game theory. Therefore, the chemical behavior and physical treats of any chemical reaction can be predicted. Multiple examples of each batch of the chemical reactions is expressed to support the useability of our claims.

\end{abstract}


\maketitle

\section{Introduction}

Game theory was for first time recognized by Von Neumann and Morgenstern (1953) in reference to human economic behavior. It is widely used in social and behavioral sciences, ranging from politics and market economics to ecosystems and biological phenomena. In game theory, a player is said to be rational if he seeks to play in a manner which maximizes his own payoff. The first explicit use of game theory terminology in the field of evolution was by Hamilton (1967), who sought for an ``{\it unbeatable strategy}\,'' for the sex ratio when there is local competition for mates. Hamilton's ``{\it unbeatable strategy}\,'' is essentially the same as an ESS as defined by Maynard Smith  and Price ( 1973). Evolutionary game theory is a way of thinking about evolution at the phenotypic level when the fitness of particular phenotypes depend on their frequencies in the population \cite{Maynard}.

Chemistry is the science that is concerned with the characterization, composition and transformations of matter \cite{Mortimer}. In $1869$, the Russian chemist Dmitri Mendeleev produced the first orderly arrangement, or periodic table, of all $63$ elements known at the time. Moseley discovered that appropriate structure of the periodic table correlated to the atomic number.
The periodic table is a very useful for correlating the properties of the elements. Molecules and ions are two important types of particles derived form atoms. Atoms are held together in molecules by chemical bonds \cite{Mortimer}.

  A central assumption of classical game theory is that the players will behave rationally, and according to some criterion of self-interest. Such an assumption would clearly be out of place in a chemical context. Instead, the criterion of rationality is replaced by that of population dynamics and stability, and the criterion of self-interest bonding formation.
By these explanations, now following questions raised as fundamental aims of the paper.
\begin{itemize}
  \item Why are doing atoms bond together?\\
  \item Why more materials in combination can be found in nature?\\
  \item Why the noble gases are to be atomic in nature?\\
\end{itemize}
Response to these questions, help us to explain the relation of the stability and the game theory. The answers must be found in the properties of atoms. Noble gases have eight electrons in their valence shell (except helium), that is the reason for their stability. To answer the first and second mentioned questions, it should point out that the atoms bond together to bring stability. Thus, the elements may acquire stability through the bonding formation.\\
Reactions of atoms depends upon many factors. The nuclear charge, the electron configuration and effective size of the atom, are the most important factors among them. Behavior of elements is determined by the distribution of electrons in around the nucleus. In other words, the electrons in the valence shell and energy levels lead to the formation of a pattern of behavior in elements. These behavior is called  ``{\it type}\,'' or ``{\it behavior kind}\,'' which are related to the  model configuration. The diversity or differences in their properties leads to heterogeneous in periodic table. Therefore, various properties observe in a period or group. The kind of behavior by any member of the population is not selective, because of specific combinations of electrons, protons and neutrons are selected naturally. Atom portion has two parts that are electrically charged, the electron has a negative charge and the nucleus has a positive charge. Atoms of each element or molecules in the reaction called a \emph{player}. The players do not make conscious choices. Interaction between players leading to bonding formation. This interaction between atoms leading to attraction force and repulsion force which these forces cause approaching atoms and bonding formation. In other words, each player achieve stable state. Noble gases has little tendency to bonding formation. It follows that they are usually stable.\\
The chemical reactions are evaluated based on \emph{collision} theory and \emph{transition state} theory. A necessary condition for both mentioned theories is collisions between particles. This work is established based on the collision theory in which the collisions between particles is formulate as a game between the elements.\\
The remainder of this paper the game of collisions is described. Chemical bond is formed when atoms combine by changes in electron distribution. There are three fundamental types of bonding:
\begin{itemize}
  \item Ionic bonding
  \item Covalent bonding
  \item Metallic bonding.
\end{itemize}
All reactions involving two reactants required collisions between particles to proceed \cite{Mortimer}. Elements can acquire stability through bonding formation. Our framework allows for implement different games for these type of reactions.
\vskip 2mm
\begin{center}
\section {\bf{Ionic bonding}}
\end{center}

An ion is a particle that is made up of an atom or a group of atoms and that bears an electric charge. There are two types:
\begin{itemize}
  \item A cation has a positive charge (because one or more electrons have been lost).
  \item An anion has a negative charge (because one or more electrons have been gained).
\end{itemize}
Ionic bonding results when electrons are transferred from type of atom to another. The atoms of one of the reacting elements lose electron and become positively charged ions. The atoms of other reactant gain electrons and become negatively charged ions. The electrostatic (plus-minus) attraction between the oppositely charged ions holds them together in a crystal structure. The net attraction holds the crystal together and may be considered to be  ionic bond \cite{Mortimer}.\\

Consider the reaction of a sodium atom and a chlorine atom. Sodium is in group $I A$. Chlorine is a member of group $VII A$. The sodium atom loses one electron; the chlorine atom gains an electron.

Consider a population of two atoms. Suppose that, this population is made up of two types, which can gain electron (G) or lose electron (L). The members of this population can not shared electrons. Is desirable to loses electrons per atom or electron gain, to achieve stable condition.

The important point of this game for particles is stability through bonding formation which has benefit and gain for both particles.
As stated earlier, we believe that each atom as a player in the position of the collisions. Each atom play one role of several kind in the population. Two players with same role and singe each earns a reward $S$ and for non-kind pair receive payment $P$. We can represent the game in a Table 1, each row of the matrix represents a kind of possible types in the population, and each column represents a possible kind of types in the population. It should be noted that player cannot choose from ``{\it lose electron}\,'' or ``{\it gain electron}\,'', however, is from kind lose electron or kind gain electron. Two players randomly chosen to play the following table:

\begin{table}[h]
\begin{tabular}{|cccc|}\cline{1-4}
\multicolumn{3}{|r} {Table 1. A symmetric game between sodium and chlorine}&\\
\multicolumn{3}{|r}{atoms with no symmetric Nash equilibrium \ \ \ \ \ \ \ \ \ \ \ \ \ \ \ \ \ \ }&\\
 & \multicolumn{1}{r}{G} \ \ \ & \ \ \ \ \ \ {L} & \\  \cline{2-3}
   G & \ \ \ \ \ \ \ \ \ \ \ \ \ \  (S, S) & \ \ \ \ \ \ (P, P) &\\
   L & \ \ \ \ \ \ \ \ \ \ \ \ \ \  (P, P) & \ \ \ \ \ \ (S, S) &\\ \cline{1-4}
\end{tabular}
\end{table}
This game has two  Nash equilibria  $(G,L)$, $(L,G)$ and ( $0\leqslant S<P$) and two Nash equilibriums  are polymorphism homogeneous of stable strategy. This game looks like playing anti coordination. Thus, the number of sodium ions produced is the same as the number of chlorine ions produced, and the formula, NaCl, give the simplest ratio of ions present in the compound ($1$ to $1$) \cite{Mortimer}.
\begin{align*}
Na^{+} + Cl^{-} \longrightarrow NaCl.
\end{align*}
In above example, the population is changed due to the new types of mutations. New population consists of two types of ions, means cation (C) and anion (A) that they called mutated. In collision any member of the new population, congener repulse each other and Non-kind attract each other. In case of congener each of them earns a reward $S$ and for the case of non-kind, then both receive a payment $P$. To map this collision to the Nash equilibria, we consider cation and anion as players in the game theory. Two players randomly chosen to play the following table:

\begin{table}[h]
\begin{tabular}{|cccc|}\cline{1-4}
\multicolumn{3}{|r} {Table 2.\ \ \ The payoff matrix game between cations and anions. }&\\
\multicolumn{3}{|r}{If $S<P$ the game has asymmetric Nash equilibrium. \ \ \ \ \ \ \ \ \ }&\\
\multicolumn{3}{|r}{If $S>P$ the game has symmetric Nash equilibrium. \ \ \ \ \ \ \ \ \ \ \ }&\\
   & \multicolumn{1}{r}{C} \ \ \  & \ \ {A} & \\  \cline{2-3}
   C & \ \ \ \ \ \ \ \ \ \ \ \ \ \ \ (S, S) & \ \ (P, P) &\\
   A & \ \ \ \ \ \ \ \ \ \ \ \ \ \ \ (P, P) & \ \ (S, S) &\\ \cline{1-4}
\end{tabular}
\end{table}
To formulate the game of the players, we consider the repulse energy of the species of congener is equal. We have two Nash equilibriums  $(C,A)$ and $(A,C)$, if $S<P$ which are polymorphism homogeneous of stable strategy. Consequently, the attraction force between non-kind players is more than the repulsion force of between congener players. This causes the crystal body to formed. Accordingly, if $S>P$, then we have two Nash equilibriums  $(C,C)$, $(A,A)$ and  two Nash equilibriums  are polymorphism homogeneous (polymorphism heterogeneous) stable strategy. Thus, the attractive force towards each the non-player type is less than the force of repulsion fellow players and consequently rupture crystal body.

\vskip 2mm
\begin{center}
\section {\bf{Covalent bonding}}
\end{center}


In covalent bonding electrons are shared, not transferred. A single covalent bond consists of a pair of electrons shared by two atoms. Molecules are made up of atoms covalently bonded to each other. A molecule is a particle that is formed from two or more atoms that are bound tightly together. The collision theory describes reactions in terms of collisions between reacting molecules (players). Assume that the hypothetical gas-phase reaction:
\begin{equation*}
A_2(g) + B_2(g) \rightarrow 2AB(g)
\end{equation*}
takes place through collisions between $A_2$ and $B_2$ molecule (players). An $A_2$ and $B_2$ molecule strike one another. The old $A-A$ and $B-B$ bonds break while simultaneously two new $A-B$ bonds form, and two $AB$ molecules leave the scene of the collision. When two slow-moving molecules approach one another closely, they rebound because of the repulsion due to the charges of their electron clouds \cite{Mortimer}.
\\
Consider the bond formed by two fluorine atoms. The fluorine molecule can be represented by the symbol $F-F$. Assume that, there exist population of two atoms.  Each atom alone can not achieve a stable situation. Atoms shared electrons to reach a stable state. Bonding pair of electrons from the nuclei of two atoms absorbed and from electrons excreted. So, two kinds of atoms compete for absorb bonding pair of electrons. Each player has two strategies that can attract and repulse the bonding pair of electrons. One $F$ atom attracts electrons to the same extent as any atoms, such as $F$. In other words, the electron cloud density of the bond is symmetrically distributed around  two nuclei. If both attract(A), then each earns a reward $S$. If one repulse (R) and the other attract, then the repulse gains a payoff $P$ and  attract gets $S$ $(S>P)$. If both repulse, then both have payment $P$. The table below show the game:

\begin{table}[h]
\begin{tabular}{|cccc|}\cline{1-4}
\multicolumn{3}{|r} {Table 3. A purely covalent bond in molecule}&\\
\multicolumn{3}{|r}{formed form two identical atoms, such as $F_2$}&\\
 & \multicolumn{1}{r}{A} \ \ \ & {R} & \\  \cline{2-3}
  A & \ \ \ \ \ \ \ \ \ \ (S, S) & (S, P) &\\
  R & \ \ \ \ \ \ \ \ \ \ (P, S) & (P, P) &\\ \cline{1-4}
\end{tabular}
\end{table}
The strictly dominates strategies for each player attract the bonding pair of electrons $(S>P)$. We see that (A,A) is the unique Nash equilibrium. Therefore, the attract is a monomorphism homogeneous stable strategy.\\

Accordingly again we consider the bond formed by two hydrogen and chlorine atoms. The  hydrogen chloride molecule can be represented by the symbol H-Cl. We consider a population of two atoms which shared electrons to reach a stable state. Each atom alone can not achieve a stable situation. Bonding pair of electrons from the nuclei of two atoms absorbed and from electrons excreted. So, two kinds of atoms compete for absorb bonding pair of electrons. Each player has two strategies, that can attract(A) and repulse(R) the bonding pair of electrons. In the HCl molecule the electrons of the covalent bond are more strongly attracted by the chlorine atom than by the hydrogen atom. Hence, the electrons cloud density of the bond is not symmetrically distributed around  two nuclei.  If player 1(hydrogen) and player 2(chlorine) attract, then each earns a reward $S$ and $P$ respectively.  If player 1 and player 2 repulse, then each earns a reward $K$ and $T$ respectively. The payoffs rank for this game is given by $K<T<S<P$. Then the payoff matrix of our game is thus given by Table 4.

\begin{table}[h]
\begin{tabular}{|cccc|}\cline{1-4}
\multicolumn{3}{|r} {Table 4. The outcomes of play between}&\\
\multicolumn{3}{|r}{hydrogen and chlorine atoms\ \ \ \ \ \ \ \ \ \ \ \ \ \ }&\\
 & \multicolumn{1}{r}{A} \ \ \ & \ \ {R} &  \\  \cline{2-3}
   A & \ \ \ \ \ \ \ \ \ \ \ (S, P) & \ \ \ (S, T) &\\
   R & \ \ \ \ \ \ \ \ \ \ \ (K, P) & \ \ \ (K, T) &\\ \cline{1-4}
\end{tabular}
\end{table}

Since for each player the action A is strictly dominates the action R, the strategy pair
(A,A) is a Nash equilibrium. Thus, the attract is a polymorphism homogeneous stable strategy.\\

Pair of members of a population engage in the following game. Atoms in a molecules can shared electrons with congener or non-kind atom in other molecule and to reach stable state. Each player can shared electrons with non-kind, we say has chosen cooperation or shared electrons with congener, has chosen non-cooperation (defect). If both cooperate(C), then each earns a reward $S$. If one defect(D) and the other cooperate, the defector, gains a payoff P, and  cooperator gets zero. If both defect, then both a payment $P$, where payoffs satisfies the inequality $0 <P <S$.

\begin{table}[h]
\begin{tabular}{|cccc|}\cline{1-4}
\multicolumn{3}{|r} {Table 5. A symmetric game between}&\\
\multicolumn{3}{|r}{atoms two molecule different\ \ \ \ \ \ \ \ \ \ \ }&\\
 & \multicolumn{1}{r}{C}\ \ \ & \ \ \ {D} &  \\ \cline{2-3}
   C & \ \ \ \ \ \ \ (S, S)& \ \ \ (0, P) &\\
   D & \ \ \ \ \ \ \ (P, 0)& \ \ \ (P, P) &\\ \cline{1-4}
\end{tabular}
\end{table}
As mentioned earlier in this case, the game has two Nash equilibriums  (C,C) and  (D,D). Accordingly, the cooperation is a polymorphism homogeneous stable strategy. Also,  the defect is a polymorphism heterogeneous (polymorphism homogeneous) stable strategy. Our game looks like the stag hunt game.\\

Under normal conditions the group of noble gases and other elements do not be reactive. The  properties of the noble gases reflect their very stable electronic configurations. Game between the molecules or atoms(except fluorine and oxygen atoms) and noble gases: \\
If player 1 and player 2 cooperate (C), then each earns a reward S. If one defects (D) and the other cooperate, the defector gets an even larger payment P, and the cooperator gets zero. However, if both defect, then both get a payment T. The payoff rank for the game Table 6 is given by $0 <S <T <P$.

\begin{table}[h]
\begin{tabular}{|cccc|}\cline{1-4}
\multicolumn{3}{|r} {Table 6. The payoff matrix game between}&\\
\multicolumn{3}{|r}{a molecule and noble gases\ \ \ \ \ \ \ \ \ \ \ \ \ \ \ \ \ \ \ \ }&\\
 & \multicolumn{1}{r}{C}\ \ \ & \ \ \ {D} & \\ \cline{2-3}
   C & \ \ \ \ \ \ \ \ \ (S, S)& \ \ \ (0, P) &\\ ‎
   D & \ \ \ \ \ \ \ \ \ (P, 0)& \ \ \ (T, T) &\\ \cline{1-4}
\end{tabular}
\end{table}
 The dominates strategy for each player is a defect. The actions pair (D,D) is a Nash equilibrium. The  defect  is a polymorphism homogeneous  (polymorphism heterogeneous) stable strategy. Our game looks like the deadlock game.\\

Games between the fluorine and the group of noble gases:

\begin{table}[h]
\begin{tabular}{|cccc|}\cline{1-4}
\multicolumn{3}{|r} {Table 7. Game between the}&\\
\multicolumn{3}{|r}{fluorine and the noble gases }&\\
  & \multicolumn{1}{r}{C}\ \ \  & \ {D} &\\ \cline{2-3}
   C & \ \ \ \ \ \ (S, S)& (0, P) &\\
   D & \ \ \ \ \ \ (P, 0)& (T, T) &\\ \cline{1-4}
\end{tabular}
\end{table}
The payoff rank for the game Table 7 satisfies the inequality  $0 <T <P <S$. We conclude that the game has two Nash equilibriums  in pure strategies, namely (C,C) and (D,D), that two Nash equilibriums  are  polymorphism homogeneous stable strategy. Our game looks like the security dilemma game.\\

 Suppose that, population is made up of two types, which can be $H_2$ molecule or $I_2$ molecule. The collision $H_2$ and $I_2$, can lead to a reaction. Collisions that produce reactions, called ``{\it effective collisions}\,''. We assume that collisions are effective collision. If two $H_2$ molecules collision, then each earns a reward zero. If an $H_2$ and $I_2$ molecule strike one another, then both a payment P $(P>0)$.  If two $I_2$ molecules collision, gains a payoff zero.

\begin{table}[h]
\begin{tabular}{|cccc|}\cline{1-4}
\multicolumn{3}{|r} {Table 8. Collision ‏$H_2$ and $I_2$ molecules}&\\
\multicolumn{3}{|r}{for iodine chloride formed\ \ \ \ \ \ \ \ \ \ \ \ \ \ \ \ \ \ }&\\
  & \multicolumn{1}{r}{$I_2$}\ \  &{$H_2$} &\\ \cline{2-3}
   $I_2$ & \ \ \ \ \ \ \ \ \ (0, 0) & (P, P) &\\
   $H_2$ & \ \ \ \ \ \ \ \ \ (P, P) & (0, 0) &\\ \cline{1-4}
\end{tabular}
\end{table}
Observe that this game has two Nash equilibriums  $(I_2,H_2)$ and $(H_2,I_2)$. Consequently, two Nash equilibriums  are polymorphism homogeneous of stable strategy.
\vskip 2mm
\begin{center}
\section {\bf{Chemical equilibria}}
\end{center}

‎

All reversible processes tend to attain a state of equilibrium. For a reversible chemical reaction, an equilibrium state is attained when the rate at which a chemical reaction is proceeding equals the rate at which the reverse reaction is proceeding. Consider a hypothetical reversible reaction
\begin{align*}
A_2 + B_2 \rightleftharpoons 2AB.
\end{align*}
This equation may be read either forward or backward. If $A_2$ and $B_2$ are mixed, they will react to produce $AB$. A sample of pure $AB$, on the other hand, will decompose to form $A_2$ and $B_2$. An equilibrium condition is a dynamic on in which two opposing changes are occurring at equal rates. At equilibrium, the concentrations of all substances are constant, because the rates of the opposing reactions are equal and not because all activities has stopped \cite{Mortimer}.
In the process of dynamic equilibrium in the apparent scale macroscopic properties are constant, but at molecular scale, microscopic properties the dynamics of the state  are occurring.
A physically distinct portion of matter that is uniform throughout composition and properties is called a phase. Homogeneous materials consist of only one ``{\it phase}\,''. Heterogeneous materials consist of more than one phase.
Stability process (namely wherever the reaction was stopped or occurring at equal rates) has three states:

\begin{enumerate}
  \item At the process of stability in the population of a type that are in a phase be formed. Thus, it called called the ``{\it monomorphism homogeneous}\,' stable strategies.
  \item At the process of stability to the population of two or more types which are in a phase be formed. So called the ``{\it polymorphism  homogeneous}\,'' stable strategies.
  \item At the process of stability  to the population of two or more types be formed which are in two or more phases. Hence, called the ``{\it polymorphism heterogeneous}\,'' stable strategies.
\end{enumerate}
Nash equilibrium in Table 3 is a monomorphism homogeneous. Nash equilibria in Tables 1,4  and 8 is a  polymorphism homogeneous.

Game theory is a powerful tool for predicting and analyzing the behavior of various phenomena in different sciences. In the present study, we find that in the stationary state, there exist games between atoms and  molecules. Our study provides evidence for a complex communication, inter elements.  Before each experiment to produce a new material through theory of games, can be wrote play between constituent elements the material, according to game theory the concluded that the experiment process can be done or not.


\end{document}